\documentclass[10pt,conference]{IEEEtran}
\pdfoutput=1 
\IEEEoverridecommandlockouts
\usepackage{cite}
\usepackage{amsmath,amssymb,amsfonts}
\usepackage{algorithmic}
\usepackage{graphicx}
\usepackage{textcomp}
\usepackage{xcolor}
\def\BibTeX{{\rm B\kern-.05em{\sc i\kern-.025em b}\kern-.08em
    T\kern-.1667em\lower.7ex\hbox{E}\kern-.125emX}}
    
\usepackage{subfigure}
\usepackage{pifont}
\usepackage{colortbl}
\usepackage{arydshln}
\usepackage{multirow}
\usepackage{multicol}
\usepackage{booktabs}
\usepackage{verbatim}
\newcommand{\xz}[1]{{\color{red}\ding{46}[XZ: #1]}}

\begin{document}

\title{PyART: Python API Recommendation in Real-Time}

\author{\IEEEauthorblockN{Xincheng He\IEEEauthorrefmark{1}, Lei Xu\IEEEauthorrefmark{1}\IEEEauthorrefmark{3}\thanks{\IEEEauthorrefmark{3} Corresponding author}, Xiangyu Zhang\IEEEauthorrefmark{2}, Rui Hao\IEEEauthorrefmark{1}, Yang Feng\IEEEauthorrefmark{1} and Baowen Xu\IEEEauthorrefmark{1}}

\IEEEauthorblockA{\IEEEauthorrefmark{1} State Key Laboratory for Novel Software Technology, Nanjing University, China} 
\IEEEauthorblockA{\IEEEauthorrefmark{2} Purdue University, USA}
\IEEEauthorblockA{xinchenghe2016@gmail.com, xlei@nju.edu.cn,  xyzhang@cs.purdue.edu,\\ rui.hao.gm@gmail.com, fengyang@nju.edu.cn, bwxu@nju.edu.cn}}


\maketitle

\begin{abstract}
API recommendation in real-time is challenging for dynamic languages like Python. 
Many existing API recommendation techniques are highly effective, but they mainly support static languages. A few Python IDEs provide API recommendation functionalities based on type inference and training on a large corpus of Python libraries and third-party libraries. As such, they may fail to recommend or make poor recommendations when type information is missing or target APIs are project-specific.
In this paper, we propose a novel approach, PyART, to recommending APIs for Python programs in real-time.
It features a light-weight analysis to derive so-called optimistic data-flow, which is neither sound nor complete, but  simulates the local data-flow information humans can derive. It extracts three kinds of features: data-flow, token similarity, and token co-occurrence, in the context of the program point where a recommendation is solicited. A predictive model is trained on these features using the Random Forest algorithm.  
Evaluation on 8 popular Python projects demonstrates that PyART can provide effective API recommendations. When historic commits can be leveraged, which is the target scenario of a state-of-the-art tool ARIREC, our average top-1 accuracy is over 50\% and average top-10  accuracy over 70\%, outperforming APIREC and Intellicode (i.e., the recommendation component in Visual Studio) by 28.48\%-39.05\% for top-1 accuracy and 24.41\%-30.49\% for top-10 accuracy. In other applications such as when historic comments are not available and cross-project recommendation, PyART also shows better overall performance.
The time to make a recommendation is less than a second on average, satisfying the real-time requirement.

\end{abstract}


\section{Introduction}\label{Introduction}

APIs are widely used in modern software development to simplify the process of software implementation and maintenance. Recently, many approaches \cite{huang2018api,xiong2018automating,rahman2016rack,sun2019enabling,qi2019finding,yuan2019api,ling2019graph,chen2019generative,nguyen2016api,nguyen2015graph,liu2018effective,nguyen2019focus,chen2019mining,rendemystify} have been proposed to provide intelligent API recommendation.  However, most existing API recommendation approaches are mainly for statically typed languages, such as Java. Few can provide effective and efficient API recommendations for dynamic languages, such as Python and JavaScript, due to the difficulties in handling their dynamic features. Python is one of the most popular programming languages\footnote{https://octoverse.github.com/}. Many popular Machine Learning and Deep Learning programming platforms such as Tensorflow and PyTorch support applications written in Python. As such, an effective API recommendation solution for Python is of importance. 

However, it is challenging to achieve the goal.
Python applications are dynamically typed.
The type of a variable is not explicitly declared and hardly known until runtime. 
Traditional type inference techniques may not be effective for Python because 
the type of a variable at a given program point
may change along different paths and/or with different inputs as Python allows changing types, attributes, and methods of objects in an arbitrary fashion. 
The lack of type information greatly degrades the accuracy of traditional static analysis on Python and increases the uncertainty of API recommendation, which heavily relies on these analyses.
For example, in order to handle path sensitivity, traditional static analysis typically merges results along different paths, even if many of them are bogus. 
As such, a variable may have a long list of types and thus
the 
IDE (Integrated Development Environment) has to provide an over-sized list of recommendations to cover all these possible types.

Besides type inference, many traditional static analyses, such as data-flow analysis and alias analysis, have difficulty handling Python too.
 Many existing API recommending approaches are built on these analyses and hence do not support Python.
 For example, data-flow features are widely used in existing API recommendation approaches~\cite{liu2018effective,d2016collective,xie2019hirec}. However, it's challenging to extract accurate data-flow for Python programs. 
 Existing Python data-flow analyses tools produce substantial false dependencies~\cite{gorbovitski2010alias,d2016collective,fritz2017cost,xu2016python,salib2004starkiller}.
 Hence, as we show in Section~\ref{sec:evaluation}, they lead to poor  recommendation accuracy.

Furthermore, providing {\em real-time} API recommendation poses additional challenges. Recommendations are solicited during development, where
the syntax and semantics of the current context of a {\em recommendation point} (i.e., an API invocation point where the developer expects the IDE to fill in) are incomplete. Hence it is challenging to perform sophisticated static analysis.
Developers often expect the IDE to make instant recommendations,  which preclude any heavy-weight online analysis. A few Python IDEs provide API recommendation functionalities.
For example, Pycharm relies on python-skeletons\footnote{https://github.com/JetBrains/python-skeletons} and typeshed\footnote{https://github.com/python/typeshed} to make API recommendations.
Visual Studio IntelliCode\footnote{https://visualstudio.microsoft.com/zh-hans/services/intellicode} leverages Machine Learning to learn programming patterns from a  huge repository of Python projects. As we will show in Section~\ref{sec:motivation}, they still have various drawbacks, caused by the aforementioned challenges.
For example,
IntelliCode focuses on learning APIs of standard libraries and popular third-party libraries, it can hardly handle APIs defined within the same project. 


A state-of-the-art API recommendation research prototype for Java is APIREC\cite{nguyen2016api}.
Given an invocation for which the developer wants API recommendation, APIREC first extracts a bag of fine-grained atomic code changes and a set of code tokens that precede the current editing location. Then it computes the likelihood score
for each candidate API by looking up the co-occurrence frequency of code changes and tokens in a large corpus. 
It heavily relies on fine-grained code changes that happen on Abstract Syntax Tree (AST) nodes and hence the accuracy of recommendation results is heavily dependent on the accuracy of AST differencing tools and the quality of code change histories. In this paper, we ported APIREC to Python and use it as a baseline.
Our results show that it is not as effective as on Java due to the inherent challenges of analyzing Python programs. 

In this paper, we propose a new real-time API recommendation approach for Python called PyART (\underline{Py}thon \underline{A}PI \underline{R}ecommendation in Real-\underline{T}ime). 
Compared to existing solutions, 
PyART can recommend both  library APIs and APIs defined within the same project; it does not rely on any third party tools; it is lightweight and provides recommendations without noticeable delay; and it delivers recommendations with good accuracy. 
The key to the success of many existing API recommendation projects lies in the use of high quality data-flow information~\cite{liu2018effective,d2016collective,xie2019hirec}. 
They largely utilize data-flow analyses in mature compilers/analysis-infrastructures, which tend to be conservative as they were designed to ensure safety in code transformation.
Such conservativeness  is substantially aggravated by the uncertainty in dynamic languages.
We observe that the conservativeness is not necessary in our context because the essence of API recommendation is to precisely model the joint-distribution of APIs and their surrounding syntactic entities and semantic properties. The semantic properties may not need to be as accurate and complete as in the traditional application scenarios (e.g., code transformation and bug finding), as long as the approximate ones and the APIs have regularity in their joint-distribution.
As such, we develop a technique to extract so-called {\em optimistic data-flow} from the context preceding a recommendation point. This data-flow is derived in a way similar to how humans reason about data-flow: {\em heavily relying on local variables, function ids, syntactic structures, and ignoring global effects such as those by aliasing.}
Specifically, our technique first collects a comprehensive list of candidate APIs, without relying on type inference tools like IntelliCode does. It then extracts and encodes three kinds of features in the context of the recommendation point: {\em optimistic data-flow}, {\em token similarity along data-flow} that measures if the data-flow paths reaching the recommendation point involve some token similar to a candidate API, and {\em token co-occurrence} that models the joint-distribution of a candidate API with a token in its neighborhood. During training, these features are used to construct a predictive model based on {\em Random Forest}. During deployment, these features are provided to the trained model to generate a ranked list of recommendations.



Our main contributions  are summarized as follows:

\begin{itemize}
\item We propose to derive optimistic data-flow that is neither sound nor complete, but sufficient for API recommendation and cost-effective to collect. 

\item We propose to use three kinds of features: data-flow, token similarity and token co-occurrence. We also develop a method to encode them to feature vectors.


\item We develop a prototype PyART based on the proposed idea, and evaluate it on 8 real-world projects. 
The evaluation results demonstrate that PyART can provide effective API recommendations. When historic commits can be leveraged, which is the target scenario of a state-of-the-art tool ARIREC, our average top-1 accuracy is over 50\% and average top-10 accuracy over 70\%, outperforming APIREC and Intellicode by 28.48\%-39.05\% for top-1 accuracy and 24.41\%-30.49\% for top-10 accuracy. In other applications such as when historic comments are not available and cross-project recommendation, PyART also shows better overall performance.

\item Our datasets and source code are available on Github\footnote{https://github.com/PYART0/PyART}.

\end{itemize}

The rest of this paper is organized as follows: the motivating example is presented in Section~\ref{sec:motivation}. The technical details of PyART are described in Section~\ref{Approach}. The  evaluation for our approach is shown in Section~\ref{sec:evaluation}. Related work and conclusions are in Section~\ref{related work} and Section~\ref{conclusion}.

\section{Motivating Example}\label{sec:motivation}
\begin{figure*}[htb]
 
  \includegraphics[width=\linewidth]{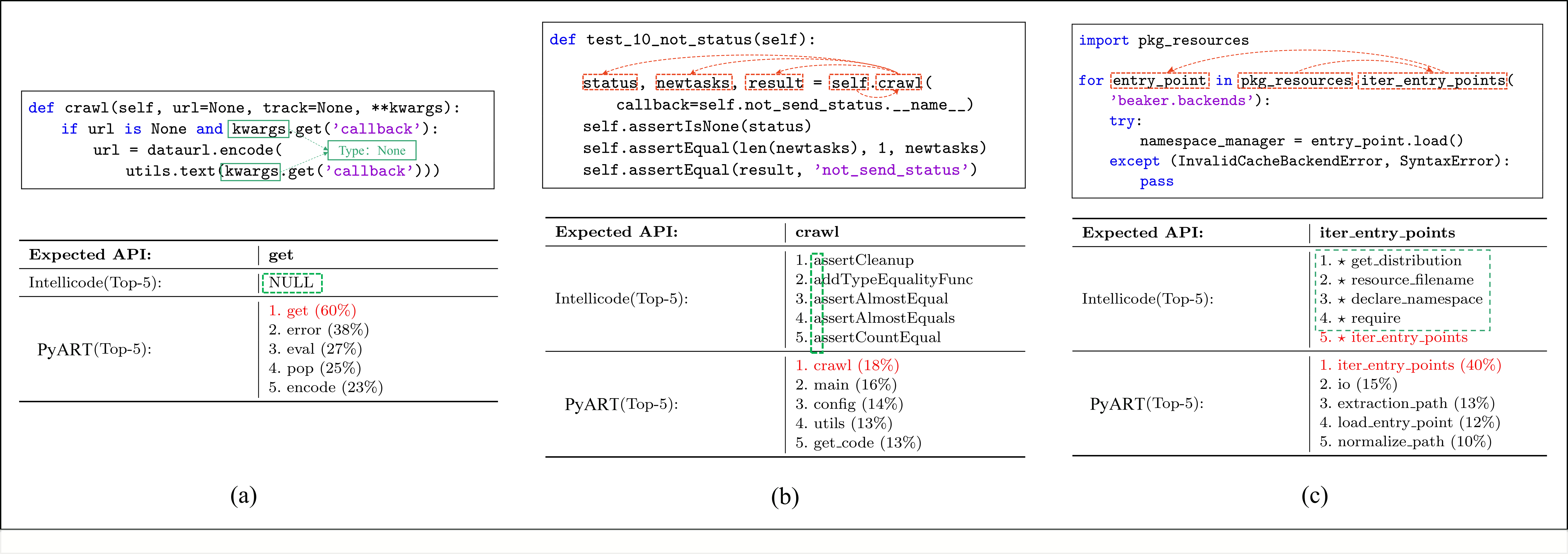}
  \vspace{-0.2in}
  \caption{Motivating example
  }\label{fig:motivation}
  \vspace{-0.2in}
\end{figure*}

API recommendation approaches based on Machine Learning tend to have better performance for Python than most traditional methods\cite{svyatkovskiy2020intellicode,svyatkovskiy2019pythia,asaduzzaman2014cscc,bruch2009learning}. However, existing Machine Learning based approaches have some limitations. We take the recommendation results of Visual Studio IntelliCode, one of the state-of-the-art recommendation tools for Python, as an example, to demonstrate these limitations (Figure~\ref{fig:motivation}).

\vspace{0.2cm}
\leftline{
\fbox{%
  \parbox{0.95\linewidth}{%
      \textit{\textbf{Observation 1.} The ability to recommend APIs is heavily dependent on the type inference result for the object of the API call.}
  }%
}}

\vspace{0.2cm}

Most recommendation tools collect candidate APIs according to the type of the object at the recommendation point. For instance, type of {\it a} in {\it a.f()}. However, due to the dynamic nature of Python, it is hard to precisely infer object type. If the type cannot be inferred, IntelliCode cannot produce any recommendation results. For example, at the recommendation point {\itshape kwargs.\_\_} in Figure~\ref{fig:motivation} (a). That is, assume the developer has keyed in ``{\it kwargs.}'' and he/she wants IntelliCode to recommend the API that is supposed to call (which shall be ``{\it get(...)}'' as shown in the final form of the code).
However, IntelliCode cannot provide any recommendation due to the lack of type information about {\itshape kwargs}.

\vspace{0.2cm}

\leftline{
\fbox{%
  \parbox{0.95\linewidth}{%
      \textit{\textbf{Observation 2.} Even if the object type can be successfully inferred, it is possible that the recommendation list contains too many candidates that do not have an order better than the alphabetical order.}
  }%
}}

\vspace{0.2cm}

The problem typically occurs at recommendation points for APIs beyond standard libraries. IntelliCode is trained based on massive data from Github to learn the most frequent programming patterns. It does very well in recommending frequently called APIs, but not the project-specific ones. For example, at the recommendation point  {\itshape  self.\_\_ } in Figure~\ref{fig:motivation} (b), the target API is one of the methods defined in a class of the current project. IntelliCode provides all of the callable methods of the class in the alphabetical order as the recommendation. Although the list contains the right answer  {\itshape crawl()}, it ranks too low to be found. The main reason is that IntelliCode is largely Machine Learning based. It is very likely that it has not seen (or trained on) many project-specific APIs. As such, its recommendation cannot be more informative than listing all callable methods.

\vspace{0.2cm}

\leftline{
\fbox{%
  \parbox{0.95\linewidth}{%
      \textit{\textbf{Observation 3.} Even if IntelliCode manages to type the object and determines a nontrivial order with some candidates labeled with '$\star$' to indicate high confidence, the recommendations can nonetheless be wrong.}
  }%
}}

\vspace{0.2cm}

 The problem occurs due to the uncertainty in Machine Learning. That is, even though IntelliCode has seen a similar API invocation context, the API that it learned may not be the one intended for the current recommendation point.
 For example, at the recommendation point {\itshape pkg\_resources.\_\_ } in Figure~\ref{fig:motivation} (c), IntelliCode provides a list of specially recommended APIs labeled by '$\star$', in which the top-4 answers are not expected but are more frequently used in the training corpus than the right API.

We propose a novel API recommendation technique specifically designed for Python. It was shown that data-flow is extremely useful in improving API recommendation results~\cite{liu2018effective,d2016collective,xie2019hirec}. However, existing techniques heavily rely on some precise data flow analysis engine, which is very difficult for a dynamic language like Python. 
Traditional data-flow analyses~\cite{kildall1973unified,rapoport2015precise,cooper2004iterative} tend to be conservative due to their application context (i.e., compilation). For example, if they cannot determine if a read and a preceding write must access different objects, they {\em conservatively} assume that there is data-flow between the two. As such, they tend to generate substantial false positives for dynamic languages where type inference and points-to analysis are much more challenging. This substantially degrades the recommendation results.
The key 
observation is {\em API recommendation can be formulated as a distribution modeling problem that does not require using conservative analysis}. In other words,
we aim to determine the probabilities of different API candidates at a recommendation point based on various kinds of hints that do not have to be sound or complete, as long as the joint-distribution of the hints and the APIs can be precisely modeled. 
Our method leverages three kinds of hints/features: {\em optimistic data-flow}, {\em token similarity along data flow}, and {\em token co-occurrence}, and uses Random Forest to learn the joint-distribution of API functions (to recommend) and these hints.
The learned model is then used to 
make recommendation.

\smallskip
\noindent
{\bf Optimistic Data-flow.}
Optimistic data-flow is data-flow that appears to be true. 
The derivation of such data-flow is not through conservative standard data-flow analysis, but rather in a way similar to how humans derive data-flow. Specifically, a set of data-flow derivation rules are defined for various syntactic structures such as function calls and loops. We say these rules are optimistic as they do not consider the possible disruption/injection of data-flow induced by aliasing. Instead, they derive data-flow that appears to be true (from variable ids and control flow structure).
For example, in Figure~\ref{fig:motivation} (b), PyART derives the optimistic data-flow at the recommendation point as {\itshape self$\rightarrow$targetAPI$\rightarrow$status$|$newtasks$|$result}.
Here, the arrows denote the data-flow direction and {\it targetAPI} denotes the API we want to predict. Observe that it only includes the data-flow that can be derived before the API invocation, to simulate the real-time API recommendation scenario where only the context before recommendation point is available.
Observe that the data-flow is optimistic as it may not be true. For example, there may not be any data-flow from {\it self} to any of the result variable. Neither is it complete. Similarly, for Figure~\ref{fig:motivation} (c), PyART extracts the optimistic data-flow {\itshape pkg\_resources$\rightarrow$targetAPI$\rightarrow$entry\_point}. 
Optimistic data-flow paths are universally extracted at all statements and linked together if possible. 
PyART then leverages the similarity of the 
data-flow paths at the recommendation point with those that it has seen in other places during training to help make prediction.


\smallskip
\noindent
\textbf{Token Similarity Along Data-flow.}
PyART also assumes that an API should have token similarity with some variable/function along an optimistic data-flow path extracted above. 
As such, for each candidate API, it measures
the similarity scores between the candidate and tokens in the data-flow paths. For example, in Figure~\ref{fig:motivation} (c), the similarity score between the target API {\itshape iter\_entry\_points} and its neighbor token in the data flow, {\itshape entry\_point}, is higher than other candidate APIs, which provides strong hint for the recommendation.

\smallskip
\noindent
{\bf Token Co-occurrence.} In the third kind of hints, PyART leverages co-occurences of tokens. That is, it predicts an API based on the tokens preceding the recommendation point (in the same source file). This includes the enclosing function, preceding variables, preceding function invocations, etc. Note that these tokens may not have data-flow with the target API. 
For example, for a recommendation point {\itshape with open(...) as f: f.\_\_ }, candidates {\itshape read()} and {\itshape write()} are more likely to be invoked due to the high co-occurrence frequency with token {\itshape open}. 
In Figure~\ref{fig:motivation} (a), PyART computes the co-occurrence of candidate API with each token in the bag: {\itshape \{def, crawl, self, url, None, track, kwargs, if\}}. And in Figure~\ref{fig:motivation} (b), PyART computes the co-occurrence of candidate API and each token in the bag {\itshape \{def, test\_10\_not\_status, status, newtasks, result, self\}}. Here, a bag is a set of tokens before the recommendation point.


These three kinds of hints are encoded as feature vectors and used to train a predictive model using Random Forest.

\begin{figure*}[htb]
 \includegraphics[width=\linewidth]{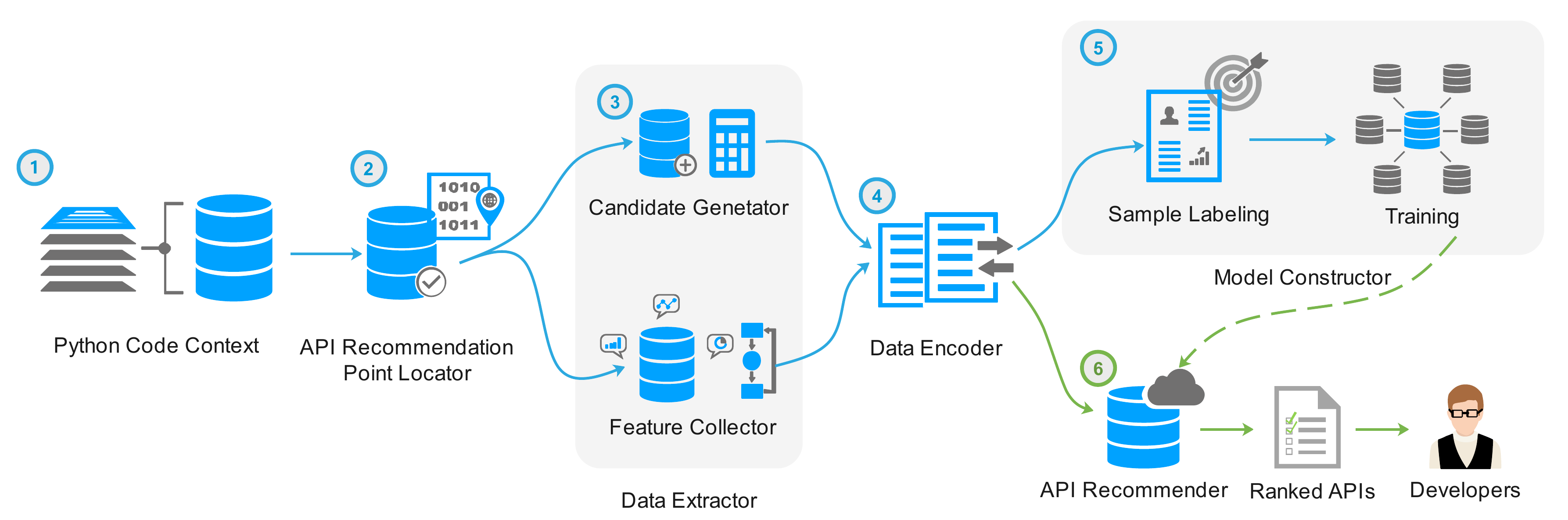}
 \vspace{-0.2in}
  \caption{Workflow of PyART
  }\label{fig:framework}
\vspace{-0.2in} 
\end{figure*}

\section{Approach}\label{Approach}

\subsection{Overview}

 The workflow of PyART is shown in Figure~\ref{fig:framework}. Basically, PyART takes  \ding{192} \emph{Python Code Context} before the recommendation point as an input, which consists of statements preceding the recommendation point (in the same file),
 and outputs a ranked list of recommended APIs to developers. It is composed of five main components: \ding{193} \emph{API Recommendation Point Locator}, \ding{194} \emph{Data Extractor}, \ding{195} \emph{Data Encoder}, \ding{196} \emph{Model Constructor} and \ding{197} \emph {API Recommender}. 

Given a python code context, the {\em API Recommendation Point Locator} first identifies the recommendation point in the form of {\itshape [Expression].targetAPI()}. The {\em Candidate Generator} in the {\em Data Extractor} first performs static type inference for the object of the target API, i.e., {\itshape [Expression]},  with an existing light-weight type inference tool pytype\footnote{https://github.com/google/pytype}. If the type inference is successful, the {\em Candidate Generator} takes all the callable methods of the inferred type as the API candidates. Otherwise, it collects API candidates from the following three sources: standard libraries, imported third-party libraries and all the callable methods declared in the current scope. As such, PyART would not yield  an empty list even if the type inference is not successful. 
For each recommendation point, the \emph{Feature Collector} in the {\em Data Extractor} 
analyzes the source code before the recommendation point to collects three kinds of hints, i.e., {\em optimistic data-flow}, {\em token similarity along data-flow} and \emph{token co-occurrence}. Then, all the hints/features are encoded to feature vectors by the Data Encoder. Each feature vector is a tuple of 4: $\vec t$ $=(t_1,t_2,t_3,t_4)$, with 
$t_1$ representing the optimistic data-flow hint, $t_2$  the token similarity along data-flow hint, and $t_3$ and $t_4$ the token co-occurrence hints, specifically, $t_3$ the co-occurrence (of an API) with the object (that invokes the API) and $t_4$ the co-occurrence with other tokens.

In model training by the {\em Model Constructor}, feature vectors are extracted from the projects in the training corpus and used to compose positive and negative samples. Positive samples are constructed with APIs and their feature vectors, and negative samples are composed of feature vectors with the API candidates other than the true positives identified by the {\em Candidate Generator}. 
Then, a real-time recommendation model is trained using the \emph{Random Forest} algorithm. It essentially learns a joint-distribution of the hints and the APIs.
In the process of API recommendation, the {\em API Recommender} 
uses the pre-trained model  to compute the probability \emph{$p_i$} of each candidate API and ranks all these candidates according to \emph{$p_i$}. 

Due to the demand of  real-time API recommendation,
these steps have to be lightweight. In the following, we discuss a number of the important steps in details.

\subsection{Feature Collector}\label{context analyzer}

\renewcommand\arraystretch{1.32}
\begin{table*}[htb]
  \caption{Optimistic Data-flow Extraction Rules}\label{tab:behaviors}
  
 \center
 { \footnotesize
  \begin{tabular}{p{4.5cm}|p{8.5cm}|p{3.5cm}}  \toprule
  Rule name & Data-flow derivation & Example \\ \hline
    $\ell:Assign(v,e)$ &$\{\forall u \in VM[\overline{\ell}](e),\  u\rightarrow v\}\subseteq DFS[\underline{\ell}](v)$ & e.g. $v = e$ \\
    \hline
    $\ell:For(v,e)$ &$\{\forall u \in VM[\overline{\ell}](e),\  u\rightarrow v\}\subseteq DFS[\underline{\ell}](v)$ & e.g. \emph{for v in e} \\
    \hline
    $\ell:Invoke(u,v)$ & $\{u\rightarrow v\}\subseteq DFS[\underline{\ell}](v)$ & e.g. $u.v$ \\
    \hline
    $\ell:Access(v,e)$ &$\{\forall u \in VM[\overline{\ell}](e),\  u\rightarrow v\}\subseteq DFS[\underline{\ell}](v)$ & e.g. $v[e]$ \\
    \hline
    $\ell:Para(f,E), E=(e_{1},\dots,e_{n})$ & $\{\forall e_{i} \in E.\forall u_{i} \epsilon VM[\overline{\ell}](e_{i}),\  u_{i}\rightarrow f\}\subseteq DFS[\underline{\ell}](f)$ &  e.g. $f(e_{1},\dots,e_{n})$\\
    \hline
    $\ell:Aggregate$, $U_i, U_j$ precedes $\ell$ & $\forall v \in U_i\cap U_j.DFS[\overline{\ell}](v)_{U_{i}} \cup DFS[\overline{\ell}](v)_{U_{j}}\subseteq DFS[\underline{\ell}](v)$ & e.g. \emph{for v in} $u.f(e,x[y])$\\
    \hline
    $\ell: Propagation$, $\ell_{1}\in pred(\ell)$ & $\forall v\in VM.DFS[\underline{\ell_{1}}](v)\subseteq DFS[\overline{\ell}](v)$ & \\
    \hline
    $\ell:Preservation$ & $\forall v \in VM \backslash kill(\ell).DFS[\overline{\ell}](v)\subseteq DFS[\underline{\ell}](v)$ & \\
     \bottomrule[.9pt]
  \end{tabular}
  }
  \vspace{-0.15in}
\end{table*}
\renewcommand\arraystretch{1.0}

Feature collection needs to address two main challenges: (1) due to the dynamic nature of Python, there are very few static analysis tools that are capable of providing accurate results for Python applications in a cost-effective manner; and (2) in the context of API recommendation, it is mandatory to deal with partial code. 
Therefore, PyART features an efficient way of extracting optimistic data-flow around the recommendation point. As mentioned in Section~\ref{sec:motivation}, such data-flow is neither sound nor complete, it just appears to be correct. The idea is to simulate how humans reason about data-flow. It is supposed to be concise and largely precise. Our hypothesis is that it is sufficient to build a joint-distribution for API prediction. Specifically, humans infer data-flow largely based on local symbolic information such as surrounding variables and program structures and rarely (and in most case unable to) reason about aliasing that is obscure and global.

Specifically, we define rules to derive optimistic data-flow from five basic {\em abstract syntax units} (AUs): assignment, loop, object attribute access/invocation, container access, and function parameter passing. As mentioned earlier, optimistic data-flow does not have to be complete such that these rules are not intended to be comprehensive but rather model the intuitive methods humans use to infer data-flow.
Table~\ref{tab:behaviors} presents these rules. In the table, label $\ell \in L$ represents the location of each unit, $ e \in E $ represents an expression, $\{u, v, x, y\} \subseteq VM$ represents a variable object or a method object, $f \in F$ represents a function in the code, and $U_i,U_j \in AU$ represent instances of the abstract syntax units. Note that Python considers any type an object, including a method.
As such, we can consider there is data-flow between an object and the API it invokes. 
The notations $\overline{\ell}$ and $\underline{\ell}$ indicate the program points immediately before and after the point labelled $\ell$, respectively. For example, $VM[\overline{\ell}](e)$ represents all the variable and method objects in expression $e$ before the location labelled $\ell$, and $DFS[\underline{\ell}](v)$ contains all the data-flow paths involving object $v$ after the location labelled $\ell$. 
The first column of the table presents the rule names; the second column presents the corresponding data-flow derivation and the last column presents examples for the corresponding AUs.

In the first rule about  assignment, for any variable and method object $u$ in the right-hand-side operand $e$ before a location $\ell$, there is (optimistic) data-flow from $u$ to the left-hand operand $v$. The second rule  $\ell:For(v,e)$ specifies data-flow extraction for a loop: there is data-flow from any variable or method object in the iterator $e$ to the loop variable $v$. The third rule $\ell:Invoke(u,v)$ is about attribute loading/invocation: if an object $u$ accesses a field attribute or invokes a method attribute $v$, there is data-flow between $u$ and $v$. Rule $\ell:Access(v,e)$ specifies data-flow for container accesses: if a container $v$, such as a list, a set or a class, is accessed through an index $e$, there is data-flow from any object $u$ in $e$ to the container object $v$. Rule $\ell:Para(f,E)$ is for function parameter passing. It specifies that any variable involved in any of the parameters $e_{1}$,\ ...,\ $e_{n}$ has data-flow to $f$. 
Since the five units may occur in combinations, rule \emph{Aggregation} aggregates the data-flow relations derived from individual units. Note that they may form paths.
Specifically, for a variable or method object $v$ occurring in both units $U_i$ and $U_j$, the data-flow of $v$ after location $\ell$ will contain both the data-flow relations generated by $U_i$ and by $U_j$ before the location $\ell$. Note that the subscripts denote the units which derive the relations. 
For example, for an expression \emph{for v in} $u.f(e,x[y])$, there are four units, including the for loop, attribute invocation, and container access and parameter passing. The feature collector combines results of each unit and outputs a long data-flow sequence: \emph{ (e$|$y$\rightarrow$x)$\rightarrow$f$\rightarrow$u$\rightarrow$v}. In addition, it models the effect of control flow following the {\em Propagation} rule: for any location $\ell$ after $\ell_1$, the data-flow relations right before $\ell$ contain all those after  $\ell_1$. The {\em Preservation} rule retains all the data-flow relations whose variable or method objects are not affected by a unit (and hence should be killed, i.e., removed). For example, the value of a global variable $a$ will not be changed by any statement in a local block when the key word $global$ is not given. In this case, data-flow relations that point to $a$ should be killed.

These rules are strictly syntax and variable name driven and do not consider aliasing. 
As such, they can be performed locally (whereas aliasing requires global analysis) and even on partial code. These are critical for real-time recommendation. According to our experiment in Section~\ref{sec:evaluation}, these rules allow us to produce data-flow with both high precision (i.e., 92.06\%-98.66\%) and high recall ( i.e., 92.58\%-98.48\%). 
In comparison, an existing analysis engine Pysonar2 has very good precision (up to 100\%) but extremely low recall (only 8.48\%-28.11\%), meaning that it can hardly be used to make API recommendations. Our results using the optimistic data-flow show that having concise and largely accurate data-flow is sufficient for achieving good recommendation outcomes.

In addition, the Feature Collector also collects the tokens involved in the data-flow relations (for the later token similarity feature encoding) and the tokens within individual source code files (for the later co-occurrence feature encoding). They are elided due to simplicity. 




\subsection{Data Encoder}\label{data encoder}

The Data Encoder
encodes the data extracted by the Data Extractor, including the API candidates and the features. As such, the encoded information can be used for training during model construction and for recommendation during deployment. 
For a candidate API, the encoder generates a feature vector $\vec t$=($t_1,t_2,t_3,t_4$) that includes the  three aforementioned features, i.e., optimistic data-flow, token similarity along data-flow and token co-occurrence.
In the following, we discuss how to encode individual features.

\smallskip
\noindent
\textbf{Optimistic Data-flow Encoding.}
Data-flow propagates important information for API recommendation.
Intuitively, assume we have two objects $a$ and $b$ and there is an assignment $b=a$ that induces data-flow between the two. According to Python semantics, $b$ inherits all the method attributes of $a$. In other words, if we want to recommend an API for $b$ (e.g., $b.$\_\_), we can get hints from what we should recommend for $a$, and vice versa. Therefore, in this part of encoding, PyART aims to quantify the similarity between the data flow path leading to the recommendation point and some paths it has seen from the corpus. 


Specifically, PyART encodes optimistic data-flow as follows. For a data-flow path $x_0\rightarrow x_1\rightarrow\dots \rightarrow \widetilde{API} \rightarrow\dots \rightarrow x_n$ that contains an API, PyART collects the corresponding token sequence $(x_0,x_1,\dots,\widetilde{API},\dots,x_n)$ and learns the order of tokens  using a common statistical language model N-Gram. With a fully pre-trained N-Gram model built on plenty of data-flow paths containing APIs, PyART is able to learn the implicit regularity between data-flow and APIs.
For instance, the N-Gram model can predict an API from the preceding data-flow sub-path. However, in PyART, the model is not directly used to make recommendation but rather to encode feature. 
During deployment, upon a recommendation request PyART encodes the features for each candidate API. It first extracts all the data-flow paths related to the current recommendation location through the Data Extractor. Since PyART focuses on real-time recommendation, we assume  it only extracts data-flow before the current recommendation point. After that, the encoder appends the API to the extracted data-flow, which includes the tokens along all the data-flow paths, sorted by their distances,
to construct an input to the trained N-Gram model. The model then outputs a log probability score \emph{p}, which is taken as the first element \emph{$t_1$} of the feature vector $\vec{t}$ of the API.

\smallskip
\noindent
{\em Example. }
For instance, given a recommendation point: \emph{for k,v in dict.\_\_}, PyART extracts two data-flow paths (ending at $k$ and $v$, respectively), denoted as $dict\rightarrow tagetAPI\rightarrow k|v$ and then appends each candidate API to the list $\{k,\ v,\ dic\}$ to construct an input to the N-Gram model. Among the inputs (from different candidates), the list denoting the true positive $dict\rightarrow items\rightarrow k|v$ 
gets the highest log probability (-3.99), higher than the probabilities of others (i.e., from -7.16 to -12.3738). The probability -3.99 is used as the first element of the feature vector for the candidate API $items$.


\smallskip
\noindent
\textbf{Encoding Token Similarity Along Data-flow.} 
Intuitively, PyART considers that a candidate API is likely the intended one if it  z~~sXCQDV1GFHWEYBSJQ1`  Q v has similarity to some token along a data-flow path reaching the recommendation point.
Specifically, for a data-flow path $x_0\rightarrow x_1\rightarrow\dots \rightarrow \widetilde{API} \rightarrow\dots \rightarrow x_n$, PyART produces a set of triples in the form of ($x_i,\widetilde{API},d$), in which $x_i$ represents a token in the path and
$d$ represents the distance between $x_i$ and $\widetilde{API}$, with  $0\leq i\leq n$,\  $0<d\leq n$. For a triple ($x_i,\widetilde{API},d$), the encoder measures the similarity  between $\widetilde{API}$ and tokevcn $x_i$. It also assigns a $x_i$ closer to $\widetilde{API}$ a larger weight (for their similarity score).
For example, in Figure~\ref{fig:motivation} (c), PyART acquires a data-flow path: {\itshape pkg\_resources$\rightarrow$targetAPI$\rightarrow$entry\_point}, in which the true positive recommendation {\itshape iter\_entry\_points()} for $targetAPI$ shares a common sub-string with the token {\itshape entry\_point}, and is thus given a larger similarity score. 

More precisely, for a token $x_i$ along some data-flow path to the recommendation point, the encoder computes the similarity score between $x_i$ and a candidate $API$ as follows.

\begin{equation}
   sim(x_i,API) = \frac{2\times|lcs_k (x_i, API)|}{d\times(|x_i| + |API|)}
\end{equation}

Note that $ lcs_k (x, API)$ computes the longest common token sub-sequence of $x_i$ and $API$.

Upon a recommendation request, the encoder  computes the total similarity score $tosim(DFS,API)$ between each candidate $API$  and the tokens in  all the data-flow paths that reach the recommendation point, denoted by a set $DFS$, as the second element $t_2$ of the feature vector $\vec{t}$ of $API$, using the following equation:
\begin{equation}
   tosim(DFS,API) = \frac{\sum_{x\in DFS} sim(x,API)}{|DFS|-1}
\end{equation}

\noindent
\textbf{Encoding Token Co-occurrence.} For each candidate API in a recommendation request, the encoder derives and encodes the token co-occurrence feature in two aspects: (1) co-occurrence between the object (whose API needs recommendation) and the candidate, called {\em object-API co-occurrence}; (2) co-occurrence between tokens in the current context (all the code up to the recommendation point) 
and the candidate (called {\em context-API co-occurrence}).
Intuitively, the first kind leverages the observation that the object and the API are the closest couple and their co-occurrence distribution provides strong indication, whereas the second kind aims to leverage a broader set of information.

For object-API co-occurrences, 
the encoder replaces the object name  with its type if available, for better generality. 
Some example type-API patterns are shown as follows: \emph{List.append()}, \emph{String.join()} and \emph{File.read()}. They are frequently used in Python projects.

More precisely, 
the encoder computes the object-API co-occurrence frequency of an object $x$ and a candidate $API$ as the third element $t_3$ of the feature vector $\vec t$ of $API$ with the following equation:

{\small
 \begin{equation}
   confidence(x\rightarrow API) = p (API|x) =
   \left\{
   \begin{array}{lr}
      \frac{N(API,x)}{N(x)}, N(x)\neq 0             \\
      \quad 0, \qquad N(x) = 0
   \end{array}
    \right.
\end{equation}
}
Here $N(x)$ represents the number of occurrences of object $x$, and $N(API,x)$ represents the number of co-occurrences (i.e., $x.API()$).

For context-API co-occurrences, the encoder considers all the tokens in the current context before the recommendation point. For example, the built-in function \emph{open()} that aims to open a file always appears with some specific tokens like \emph{with}, \emph{as}, \emph{f}, since the statement is in the form of \emph{with open(...) as f:} is widely used in software development. 
In particular, the encoder computes the co-occurrence frequency of a token $x$ and a candidate $API$ with the following equation.

\vspace{-0.1in}
{\small
\begin{equation}
   confidence(x \rightarrow API) = p (API|x) =
   \left\{
   \begin{array}{lr}
      \frac{N(API,x)}{N(x)}, N(x)\neq 0             \\
      \quad 0, \qquad N(x) = 0
   \end{array}
    \right.
\end{equation}
}
Here $N(x)$ represents the number of files in which $x$ occurs, and $N(API,x)$ represents the number of files in which $x$ and $API$ co-occur.

For a set of tokens $S$
(of the current context before the recommendation point) $S=\{x_0,x_1,\dots,x_n\}$, the encoder computes the overall co-occurrence frequency of $S$ and a candidate \emph{API} as the fourth element $t_4$ of the feature vector $\vec t$ with the following equation.
\begin{equation}
totalCnfd(S,API)=\frac{1}{|S|} \sum_{x_i}\frac{confidence(x_i\rightarrow API)}{dist(x_i,API)}
\end{equation}
Here $dist(x_i,API)$ computes the distance between $x_i$ and $API$. Recall  a closer $x_i$ has
a larger weight for their co-occurrence score.


\subsection{Training and Recommendation}\label{model}

\noindent
\textbf{Model Constructor.} The Model Constructor performs supervised learning using Random Forest. The model learns to predict API from the feature vectors. The training feature vectors are divided into two categories: (1) positive cases: feature vectors generated from the true positive APIs, labeled as \emph{1}; and (2) negative cases: feature vectors generated from other candidate APIs, labeled as \emph{0}. 

\smallskip
\noindent
\textbf{API Recommender.} For a recommendation request {\itshape [Expression].\_\_},  
PyART first collects all the candidate APIs.
For each candidate API, it computes the corresponding feature vector.
Then, the API recommender takes all the candidate feature vectors as test cases and provides them to the recommendation model that is pre-trained.
For each candidate vector, the recommendation model produces a probability $p_i$ that the label of the vector is \emph{1}. Finally, it ranks all the candidates according to $p_i$ and outputs the ranked list to developers.

\section{Evaluation}\label{sec:evaluation}
\subsection{Research Questions}
To evaluate PyART, we propose the following research questions.

\textbf{RQ1:} How effective is PyART in data-flow analysis, compared with one of the state-of-art approaches, Pysonar2?

\textbf{RQ2:} In comparison with the state-of-the-art approaches, Visual Studio IntelliCode and Py-APIREC, how effective is PyART in recommending APIs within a project?

\textbf{RQ3:} In comparison with the state-of-the-art approaches, Visual Studio IntelliCode and Py-APIREC, how effective is PyART in recommending API calls across projects?

\textbf{RQ4:} How efficient is PyART in recommending APIs in real-time?

\textbf{RQ5:} What is the impact of each kind of features?

\subsection{Baselines}

To evaluate the effectiveness and efficiency of PyART, we choose a state-of-the-art,  Pysonar2, as the baseline for data-flow analysis evaluation (RQ1), and two state-of-the-art approaches Visual Studio IntelliCode and Py-APIREC as the baselines for the API recommendation evaluation (RQ2-3).

\noindent
\textbf{Pysonar2.} Pysonar2\footnote{https://github.com/yinwang0/pysonar2} is an advanced static analyzer for Python, which performs (costly) whole-project inter-procedural analysis to infer types.

\noindent
\textbf{Visual Studio IntelliCode.} Visual Studio IntelliCode is an experimental set of AI-assisted developer productivity tools, trained from thousands of open-source projects on GitHub with high star ratings\footnote{https://visualstudio.microsoft.com}. For an API recommendation task, IntelliCode places the most relevant ones at the top of the recommendation list and labels them with a star.

\noindent
\textbf{Py-APIREC.} We reproduce a state-of-art API recommendation method for Java named APIREC\cite{nguyen2016api} and port it to Python.
Since Py-APIREC relies heavily on fine-grained code changes that happen on AST nodes, the accuracy of recommendation results is hence heavily dependent on the accuracy of AST differencing tools and code change histories. 
Following the original paper, we diff the consecutive commits to identify code changes.
We validate the correctness of the reproduced APIREC (before porting it to Python) by comparing the results acquired by our reproduced system with the ones published in the original paper. The error is within 5\%. 

\subsection{Implementation}
We make our evaluation on Linux Ubuntu 4.15.0-66-generic with  Intel(R) Xeon(R) Gold 5118 CPU @ 2.30GHz. All approaches (including Py-APIREC and PyART) are realized in Python 3. 
In the process of porting APIREC, We use the state-of-the-art AST diff tool Gumtree\footnote{https://github.com/GumTreeDiff/gumtree} to extract fine-grained code changes. We learn two parameters $w_C$ and $w_T$ for Py-APIREC from a training set  using the hill-climbing adaptive learning algorithm \cite{nguyen2016api}, and use the following settings:
(1) we set the value of fold number \emph{k} as 10;
(2) we set the step size $\delta$ in hill-climbing as 0.01;
(3) we set the maximum number of iterations in adaptive learning as 1000.
Besides, in order to provide a simple static type inference for PyART, we use a type inference tool Pytype\footnote{https://github.com/google/pytype}.

\subsection{Evaluation Setup}\label{data}

For fair comparison with Py-APIREC and Intellicode, 
we set up three scenarios about the corpora used in evaluation.

\smallskip
\noindent
\textbf{Project Edition (PE).}
We randomly collect 8 Python projects from Github that have a long development history with 12,194 commits and 1,195,994 files in total as a corpus called \emph{Project Edition}. \emph{Project Edition} was defined in \cite{nguyen2016api}, which uses the oldest 90\% of project commits for training and the 10\% most recent commits for recommendation.
In particular, we train Py-APIREC and PyART on the first 90\% commits of each project, and test Py-APIREC, IntelliCode and PyART on the remaining 10\% commits.
Since Py-APIREC can only recommend APIs in specific kinds of code changes (i.e., {\it Add} that inserts a node into an AST,  {\em MethodInvocation} that  changes a method invocation, and {\em APIName} that denotes the name of a method) and cannot deal with APIs in other kinds of changes, 
we select code changes that Py-APIREC can identify and recommend in the remaining 10\% commits as the recommendation points for a fair comparison.
The results of PE are used to answer RQ2 (intra-project effectiveness). Moreover, we record the time of each recommendation of PyART  to answer RQ4. 

\smallskip
\noindent
\textbf{Intra-Project Edition (IPE).} To evaluate the intra-project  effectiveness of PyART when historic commits are not available,  we prepare another set as follows. We call it the 
\emph{Intra-Project Edition} (IPE).
Specifically, we collect the last commit of each project in PE to form the IPE corpus, consisting of 474 Python files and 86,853 LOCs. We then divide the source files of each project in 10 folds. We train PyART on 9 of the 10 folds and test it on the 1. The 10-fold evaluation is performed five times on each project. The average results are shown in Table~\ref{tab:ipe} as part of the answer to RQ2. 

\smallskip
\noindent
\textbf{Community Edition (CE).} APIREC collects a
large corpus called {\em Community Edition}. It trains its prediction model on this corpus ad then tests it on a different (and smaller) corpus to evaluate the cross-project recommendation effectiveness. 
We conduct a similar experiment.
We collect the top-30 forked Python projects from Github, with long development histories (i.e., 86,078 commits and 7,634,717 files) as the CE corpus. We train Py-APIREC on all the commits of the 30 projects and train PyART only on the last commit of the 30 projects for the sake of  cost-saving. Then, we use Py-APIREC, IntelliCode and PyART to make recommendations for code changes in all the commits of the 8 projects in PE to answer RQ3. In addition, we train PyART with different subsets of feature vectors from CE and evaluate the trained model on the projects from PE to study how the features impact accuracy (RQ5).


The statistics for the PE and CE sets are shown in Table~\ref{table:data}.

\renewcommand\arraystretch{1.31}
\begin{table}[tb]
  \caption{Data set statistics for PE and CE}\label{table:data}
  \begin{tabular*}{\hsize}{@{}@{\extracolsep{\fill}}lll@{}}
   \toprule
     & PE& CE\\
    \hline
    Total projects & 8 & 30\\
    Total source files  &1,195,994 & 7,634,717\\
    Total SLOCs  & 166,146,606 & 1,344,587,261\\
    \hline
    Number of commits & 12,194 & 86,078\\
    Total changed files &20,390  & 106,817\\
    Total AST nodes of changed files & 7,538,521 & 113,657,525\\
    \hline
    Total changed AST nodes &1,024,072 & 10,043,518\\
    Total detected changes & 264,683 & 2,040,272\\
    Total detected changes with APIs & 21,795 & 285,216\\
    \bottomrule
  \end{tabular*}
  \vspace{-0.2in}
\end{table}
\renewcommand\arraystretch{1.0}

\subsection{Metrics and Settings}

\smallskip
\noindent
\textbf{Data-flow analysis evaluation.} We evaluate the effectiveness of data-flow analysis in PyART and also in Pysonar2 (for comparison) using precision, recall and F1-score. 
In order to construct a good baseline like the ``ground truth'', two authors that have more than three years of Python
coding experience inspect a subset of source files in IPE (10 random files per project with each file having 27-1725 LOC) and manually recognize the data-flow in these files.
Here we look at pair-wise data-flow relations (i.e., definition and use).
We use manual inspection as the ground truth because there is not an existing tool that can report sound and complete data-flow information.
We do recognize that humans may not extract global data-flow and hence the human results may be biased. Note that {\em we do not claim that our data-flow analysis is accurate and complete. Instead, optimistic data flow analysis is to simulate how humans reason about data flow and hence we argue the human study can serve as a reasonable baseline.}


\noindent
\textbf{API Recommendation evaluation.}   We evaluate PyART, Py-APIREC and IntelliCode using \emph{Top-k} accuracy(\%). For a list of possible APIs recommended with the length of \emph{l}, we search for the correct answer among the first \emph{k} elements of the list. We set the value of \emph{k} to 1, 2, 3, 4, 5, and 10, respectively. For a more intuitive and objective evaluation of PyART and the other baselines, we also use MRR as an evaluation measure. MRR (Mean Reciprocal Rank) evaluates the process that produces a list of possible APIs ordered by the probability of correctness, and computes the average of the reciprocal ranks of results with Equation~(\ref{mrr}):
\begin{equation}\label{mrr}
   MRR = \frac{1}{|Q|}\sum^{|Q|}_{i=1}\frac{1}{rank_i}
\end{equation}
$Q$ refers to a sample of queries, and $rank_i$ refers to the rank of the first relevant API for the \emph{i}-th query. Intuitively, a larger MRR value means a more accurate recommendation.

\subsection{Result Analysis}

\renewcommand\arraystretch{1.31}
\begin{table*}[htp]
  \centering
  \scriptsize
  \caption{Data flow analysis results for Python ({\bf compared to the human extracted baseline})}\label{tab:dataflow}
    \begin{tabular}{p{1.4cm} p{1.4cm} p{1.2cm} <{\centering}p{1.2cm}<{\centering}p{1.2cm}<{\centering}p{1.2cm}<{\centering}p{1.2cm}<{\centering}p{1.2cm}<{\centering} p{1.2cm}<{\centering} p{1.2cm}<{\centering}}
   \toprule
         Model & Metrics&  cornice  & pyspider  & bs4  & httpbin  & allennlp  & gitsome & simplejson & flask \\
    \hline
      \multirow{3}{*}{Pysonar2} &Precision & 100 & 100 & 100 & 100 & \textcolor[rgb]{1.00,0.00,0.00}{Failure} & \textcolor[rgb]{1.00,0.00,0.00}{Failure} & 100 & \textcolor[rgb]{1.00,0.00,0.00}{Failure}  \\
     &Recall & 9.03 & 8.48 & 14.77 & 9.83  &\textcolor[rgb]{1.00,0.00,0.00}{Failure} &\textcolor[rgb]{1.00,0.00,0.00}{Failure} &28.11 & \textcolor[rgb]{1.00,0.00,0.00}{Failure} \\ 
     &F1-score & 16.56 & 15.63 & 25.74 & 17.90 &\textcolor[rgb]{1.00,0.00,0.00}{Failure} &\textcolor[rgb]{1.00,0.00,0.00}{Failure} & 43.88 &\textcolor[rgb]{1.00,0.00,0.00}{Failure} \\
     \hline
     \multirow{3}{*}{PyART} &Precision & 96.46 & 98.54 & 95.57 & 96.75 & 95.85 & 92.06 & 97.53 & 98.66 \\
     &Recall & 94.06 & 96.97 & 98.48 & 96.65 & 92.58 & 98.03 & 96.42 & 96.00 \\ 
     &F1-score & \textbf{95.24} & \textbf{97.75} & \textbf{97.00} &  \textbf{96.70} & \textbf{94.19} & \textbf{94.95} & \textbf{96.97} & \textbf{97.31} \\
    \bottomrule
    \end{tabular}%
\end{table*}%
\renewcommand\arraystretch{1.0}

\renewcommand\arraystretch{1.31}
\begin{table}[htp]
  \centering
  \scriptsize
  \caption{Recommendaiton results within projects}\label{tab:inproj}
    \begin{tabular}{p{.95cm}|p{1.25cm}p{.4cm}p{.4cm}p{.4cm}p{.4cm}p{.4cm}p{.4cm} p{.4cm}}
   \toprule
          & Model & top1  & top2  & top3  & top4  & top5  & top10 & MMR \\
    \hline
     \multirow{3}[2]{*}{allennlp} & Py-APIREC & 17.39 & \textbf{26.09} & 26.09 & \textbf{30.43} & \textbf{30.43} & \textbf{39.13} & 23.74  \\
          & IntelliCode & 1.64 & 3.28 & 3.28 & 3.28 & 14.75 & 21.31 & 8.14 \\
         & \textbf{PyART}  & \textbf{19.70} & 23.96 & \textbf{26.46} & 28.77 & 30.06 & 34.51 & \textbf{24.95}   \\
    \hline
    \multirow{3}[2]{*}{bs4} & Py-APIREC & 27.27 & 27.27 & 27.27 & 27.27 & 27.27 & 45.45 & 31.42  \\
          & IntelliCode & 41.67 & 50.00 & 50.00 & 50.00  & 50.00 & 58.33 & 48.76  \\
          & \textbf{PyART} & \textbf{70.49} & \textbf{77.97} & \textbf{80.08} & \textbf{81.60} & \textbf{83.21} & \textbf{86.26} & \textbf{76.22} \\
    \hline
        \multirow{3}[2]{*}{cornice} & Py-APIREC & 7.50   & 15.00    & 22.50  & 27.50  & 27.50  & 32.50  & 17.04 \\
          & IntelliCode & 13.33 & 13.33 & 13.33 & 13.33 & 20.00    & 26.67 & 17.62 \\
          & \textbf{PyART} & \textbf{48.55} & \textbf{56.41} & \textbf{63.05} & \textbf{65.01} & \textbf{66.90} & \textbf{70.94} & \textbf{56.89}      \\
    \hline
    \multirow{3}[2]{*}{flask} & Py-APIREC & 5.56 & 11.11 & 11.11 & 11.11 & 16.67 & 27.78 & 12.04 \\
          & IntelliCode & 44.44 & 48.15 &48.15 & 48.15 & 48.15 & 59.26 & 48.71 \\
          & \textbf{PyART}  & \textbf{53.84} & \textbf{63.68} & \textbf{67.64} & \textbf{70.17} & \textbf{71.89} & \textbf{76.21} & \textbf{62.02}   \\
    \hline
    \multirow{3}[2]{*}{gitsome} & Py-APIREC & 4.04 & 8.31 & 10.56 & 11.01 & 11.69 & 15.06 & 8.71  \\
          & IntelliCode & 19.78 & 21.98 & 25.27 & 30.77 & 34.07 & 36.26 & 24.71 \\
          & \textbf{PyART} & \textbf{30.85} & \textbf{40.69} & \textbf{44.56} & \textbf{47.54} & \textbf{49.78} & \textbf{55.74} & \textbf{39.75} \\
    \hline
    
    \multirow{3}[2]{*}{httpbin} & Py-APIREC &11.11  & 11.11 & 22.23 & 44.44 & 44.44 & 44.44 & 24.13 \\
          & IntelliCode & 38.10 & 38.10 & 42.86 & 42.86 & 42.86 & 42.86 & 41.39 \\
          & \textbf{PyART} & \textbf{67.06} & \textbf{79.22} & \textbf{80.78} & \textbf{80.78} & \textbf{82.35} & \textbf{83.92} & \textbf{74.42} \\
    \hline
    \multirow{3}[2]{*}{pyspider} & Py-APIREC & 13.59 & 33.7 & 42.93 & 43.48 & 44.56 & 45.65 & 27.51 \\
          & IntelliCode & 6.06 & 10.61 & 13.64 & 16.67 & 16.67 & 57.58 & 15.48 \\
          & \textbf{PyART} &  \textbf{60.41} & \textbf{71.87} & \textbf{75.47} & \textbf{77.46} & \textbf{78.61} & \textbf{82.90} & \textbf{68.99}  \\
    \hline
    \multirow{3}[2]{*}{simplejson} & Py-APIREC & 8.33 & 8.33 & 16.67 & 66.67 & 75.00 & 75.00 & 26.20 \\
          & IntelliCode & 14.29 & 28.57 & 28.57 & 57.14 & 57.14 & 71.43 & 31.35  \\
          & \textbf{PyART} & \textbf{65.11} & \textbf{80.65} & \textbf{81.58} & \textbf{85.77} & \textbf{85.91} & \textbf{87.18} & \textbf{74.57}  \\
    \hline
    \multirow{3}[2]{*}{\textit{\textbf{avg}}} & Py-APIREC &  11.84     &  17.62     &  22.42     & 32.74      &    34.70   &  40.63     & 21.35  \\
          & IntelliCode &   22.41    &  26.75     &     28.14  &    32.78   &    35.46   &  46.71     & 29.52 \\
          & \textbf{PyART}  & \textbf{50.89}      &  \textbf{60.27}     & \textbf{63.43}      &    \textbf{65.54}   &  \textbf{66.99}     &  \textbf{71.12}     &  \textbf{58.47}\\
    \bottomrule
    \end{tabular}%
  \label{rq2}%
\end{table}%
\renewcommand\arraystretch{1.0}

\noindent
\textbf{Data Flow Analysis Evaluation (RQ1).} According to Table~\ref{tab:dataflow}, PyART's data-flow results align well with the human baseline (92.06\%-98.66\% precision and
92.58\%-98.48\% recall). 
In contrast, PySonar2 has very good precision (up to 100\%) but extremely low recall (only 8.48\%-28.11\%). The reason is that PySonar2 aims to avoid false dependences caused by aliasing. As such, when it is not sure, it does not report. Note that traditional data-flow analyses conservatively assume there is data-flow when they are not sure (for safety). Such sparse data-flow information generated by PySonar2 can hardly be used to make API recommendations.
Observe it has runtime failures during analysis in a few cases too. 

\smallskip
\noindent
\textbf{Effectiveness within Projects (RQ2).} According to Table~\ref{tab:inproj},  PyART achieves better accuracy and  MRR score than the baselines Py-APIREC and IntelliCode on almost all projects. The top1 accuracy of PyART is generally high (i.e., up to 70.49\%), which supports the effectiveness of PyART on recommending APIs within projects. 
In addition, we conduct an experiment using the IPE set (i.e., no historic commits).
We train the model on 90\% of the source files and evaluate it on the remaining 10\% files. The recommendation points are selected in the same way as before. 
Note that Py-APIREC is not applicable in this scenario.
 The results are shown in Table~\ref{tab:ipe}. 
 Observe that PyART is still highly effective even it does not get to learn from other commits.
 It substantially outperforms IntelliCode in majority of the cases. 

\renewcommand\arraystretch{1.31}
\begin{table}[htb]
  \centering
  \scriptsize
  \caption{Results of Intra-Project Evaluation}\label{tab:ipe}
    \begin{tabular}{p{1cm}|p{1cm}p{.45cm}p{.45cm}p{.45cm}p{.45cm}p{.45cm}p{.45cm} p{.45cm}}
    \toprule
     &  Model & top1 & top2 & top3 & top4 & top5 & top10 & MRR \\
    \hline
     \multirow{2}[2]{*}{allennlp} & IntelliCode & \textbf{13.74} & \textbf{16.44} & 18.02 & 18.24 & 20.05 & 22.07 & 17.24  \\
     & \textbf{PyART} & 13.17 & 16.01 & \textbf{19.57} & \textbf{19.93} & \textbf{20.64} & \textbf{24.91} & \textbf{17.42} \\
     \hline
     \multirow{2}[2]{*}{bs4} & IntelliCode & 23.08 & 25.00 & 25.00 & 30.77 & 44.23 & 50.00 & 29.59 \\
    & \textbf{PyART}  & \textbf{39.29} & \textbf{39.29} & \textbf{42.86} & \textbf{42.86} & \textbf{50.00} & \textbf{57.14} & \textbf{43.16} \\
    \hline
     \multirow{2}[2]{*}{cornice} & IntelliCode & 4.65 & 16.28 & 16.28 & 20.93 & 25.58 & 32.56 & 14.20  \\
     & \textbf{PyART} & \textbf{32.05} & \textbf{39.07} & \textbf{39.87} & \textbf{39.87} & \textbf{39.87} & \textbf{46.09} & \textbf{37.12}\\
     \hline
    \multirow{2}[2]{*}{ flask} & IntelliCode & \textbf{15.69} & \textbf{21.57} & \textbf{23.53} & \textbf{29.41} & \textbf{31.37} & \textbf{45.10} & \textbf{24.20} \\
    & \textbf{PyART}  & 15.18 & 17.31 & 19.44 & 19.44 & 23.80 &28.06 & 19.56\\
    \hline
   \multirow{2}[2]{*}{ gitsome} & IntelliCode & 6.71 & 8.05 & 8.05 & 8.72 & 8.72 & 10.74 & 8.72 \\
   & \textbf{PyART} & \textbf{32.30} & \textbf{36.77} & \textbf{46.05} & \textbf{48.45} & \textbf{50.52} & \textbf{56.01} & \textbf{40.02} \\
    \hline
   \multirow{2}[2]{*}{ httpbin} & IntelliCode & \textbf{15.66} & 16.87 &\textbf{26.51} & \textbf{27.71} & 30.12 & 33.73 & \textbf{21.64} \\
   & \textbf{PyART} & 11.11 &\textbf{20.64} & 20.64 & 24.60 & \textbf{30.16} & \textbf{40.48} & 20.28 \\
    \hline
   \multirow{2}[2]{*}{ pyspider} & IntelliCode & 22.75 & 28.74 & 29.94 & 30.54 & 32.34 & 37.13 & 28.18 \\
  & \textbf{PyART}  & \textbf{36.55} & \textbf{42.26} & \textbf{43.79} & \textbf{45.09} & \textbf{45.50} & \textbf{46.77} & \textbf{40.97} \\
   \hline
   \multirow{2}[2]{*}{ simplejson} & IntelliCode & 14.81 & 18.52 & 18.52 & 22.22 & 29.63 & 33.33 & 21.23 \\
  & \textbf{PyART}  & \textbf{16.05} & \textbf{30.57} & \textbf{34.91} & \textbf{36.38} & \textbf{36.38} & \textbf{41.50} & \textbf{26.53} \\
   \hline
   \multirow{2}[2]{*}{ \textbf{avg}} & IntelliCode & 14.64 & 18.93 & 20.73 & 23.57 & 27.78 & 33.08 & 20.63 \\
   & \textbf{PyART}  &\textbf{24.46} 
    &  \textbf{30.24} &  \textbf{33.39} &  \textbf{34.58} &  \textbf{37.11} &  \textbf{42.62} &  \textbf{30.63}  \\
     \bottomrule
    \end{tabular}%
\end{table}%
\renewcommand\arraystretch{1.0}

\smallskip
\noindent
\textbf{Effectiveness Evaluation across Projects (RQ3).} According to  Table~\ref{tab:acrossproj}, the average accuracy and MRR of PyART are higher than those of Py-APIREC and IntelliCode, which supports the effectiveness of PyART across projects. Although the top-k accuracies and MRR of PyART are generally higher, a small part of projects show inferior results (than Intellicode). This is acceptable since Intellicode is trained based on thousands of Python code on Github, while PyART is only trained on 30 projects.

\renewcommand\arraystretch{1.31}
\begin{table}[tb]
\label{rq3}
  \centering
  \scriptsize
  \caption{Recommendaiton results across projects}\label{tab:acrossproj}
    \begin{tabular}{p{.95cm}|p{1.25cm}p{.4cm}p{.4cm}p{.4cm}p{.4cm}p{.4cm}p{.4cm} p{.4cm}}
   \toprule
          & Model & top1  & top2  & top3  & top4  & top5  & top10 & MMR \\
    \hline
 \multirow{3}[2]{*}{allennlp} & Py-APIREC & 0 & 1.78 & 4.32 & 5.04 & 5.04 & 10.07 & 4.27  \\
          & IntelliCode & 12.32 & 18.01 & \textbf{23.38} & \textbf{30.02} & 31.12 & \textbf{46.45} & \textbf{22.28} \\
          & \textbf{PyART} & \textbf{13.59} & \textbf{18.26} & 22.05 & 24.72 & \textbf{32.29} & 45.66 & 22.09 \\
    \hline
    \multirow{3}[2]{*}{bs4} & Py-APIREC & 3.03 & 3.03 & 3.23 & 3.23 & 3.23 & 6.45 & 5.83 \\
          & IntelliCode & 11.39 & 14.77 & 15.19 & 19.41 & 26.16 & \textbf{40.93} & 18.72  \\
          & \textbf{PyART} & \textbf{13.65} & \textbf{22.60} & \textbf{23.03} & \textbf{25.59} & \textbf{29.64} & 35.39 & \textbf{21.48} \\
    \hline
        \multirow{3}[2]{*}{cornice} & Py-APIREC & 3.42 & 9.22 & 16.99 & 17.48 & 18.93 & 24.76 & 11.25  \\
          & IntelliCode & 14.00 & 19.20 & 20.40 &22.80 & 25.60 & 35.60 & 20.45 \\
          & \textbf{PyART} & \textbf{17.89} & \textbf{24.31} & \textbf{26.61} & \textbf{27.98} & \textbf{32.11} & \textbf{38.99} & \textbf{24.89} \\
    \hline
    \multirow{3}[2]{*}{flask} & Py-APIREC & 4.65 & 10.47 & 13.95 & 17.44 & 17.44 & 18.60 & 11.13  \\
          & IntelliCode & \textbf{20.79} & \textbf{25.14} & \textbf{27.98} & 29.87 & 30.25 & 37.62 & \textbf{26.38} \\
          & \textbf{PyART} &12.80 & 21.39 & 24.68 & \textbf{33.64} & \textbf{37.48} & \textbf{44.79} & 22.90 \\
    \hline
    \multirow{3}[2]{*}{gitsome} & Py-APIREC & 12.24 & 14.29 & 15.31 & 17.35 & 18.37 & 21.43 & 15.94\\
          & IntelliCode & 19.47 & 24.34 & 28.76 & 31.86 & 33.19 & 33.19 & 25.88 \\
          & \textbf{PyART} & \textbf{27.81} & \textbf{28.48} & \textbf{32.45} & \textbf{33.77} & \textbf{34.44} & \textbf{39.74} & \textbf{31.44}\\
    \hline
    
    \multirow{3}[2]{*}{httpbin} & Py-APIREC &10.77  & 20.77 & 24.62 & 26.92 & 27.69 & 48.46 &22.63 \\
          & IntelliCode & \textbf{27.19} & 30.70 & 31.58 & 31.58 & 32.46 & 35.96 & 31.19 \\
          & \textbf{PyART} & 27.16 & \textbf{32.10} & \textbf{37.04} & \textbf{39.51} & \textbf{45.68} & \textbf{53.09} & \textbf{34.95} \\
    \hline
    \multirow{3}[2]{*}{pyspider} & Py-APIREC & \textbf{16.95} & 16.95 & 18.64 & 18.64 & 22.03 & 22.03 & 18.87  \\
          & IntelliCode & 15.70 & 21.94 & 24.09 & 25.38 & 26.45 & 40.00 & 22.92 \\
          & \textbf{PyART} &15.74 & \textbf{25.63} & \textbf{30.46} & \textbf{32.23} & \textbf{35.03} & \textbf{42.13} & \textbf{24.97}  \\
    \hline
    \multirow{3}[2]{*}{simplejson} & Py-APIREC & 3.23 & 6.45 & 12.90 & 19.35 & 19.35 & 29.03 & 12.31 \\
          & IntelliCode & 20.00 & \textbf{27.50} & 27.50 & 32.50 & 35.00 & 45.00 & 28.36  \\
          & \textbf{PyART} & \textbf{26.32} & 26.32 & \textbf{36.84}  & \textbf{36.84} &  \textbf{63.16} & \textbf{78.95} & \textbf{38.08} \\
    \hline
    \multirow{3}[2]{*}{\textit{\textbf{avg}}} & Py-APIREC & 6.79 & 10.37 & 13.75 & 15.68 & 16.51 & 22.60 & 12.74   \\
          & IntelliCode & 17.61 & 22.70 & 23.99 & 27.93 & 30.03 & 39.34 & 24.52 \\
          & \textbf{PyART} & \textbf{19.37} & \textbf{24.89} & \textbf{29.15} &  \textbf{31.79} & \textbf{38.73} & \textbf{47.34} & \textbf{27.60}   \\
   \bottomrule
    \end{tabular}%
\end{table}%
\renewcommand\arraystretch{1.0}

\begin{figure}[htb]

  \includegraphics[width=3in]{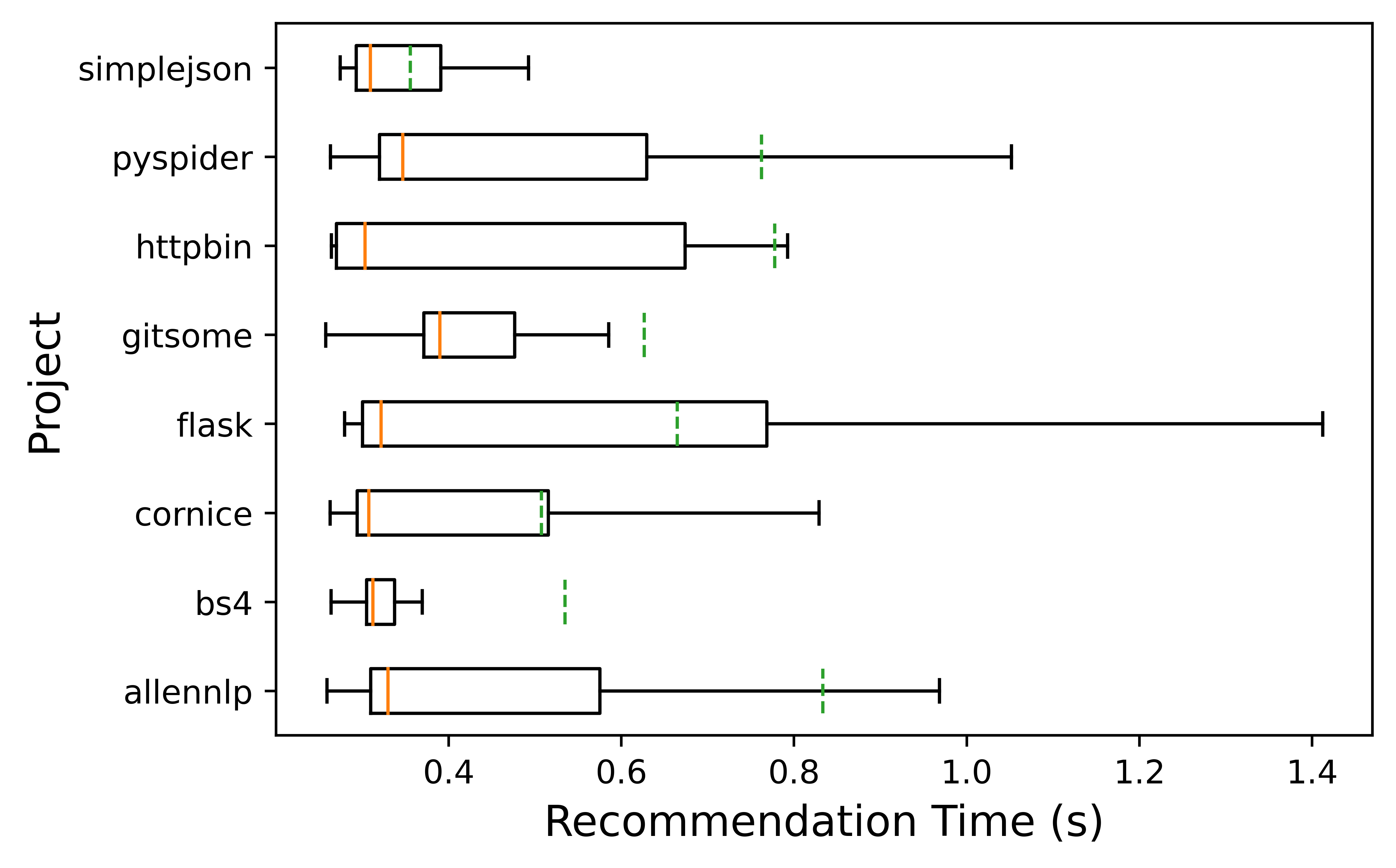}
  \vspace{-0.1in}
  \caption{Recommendation time of PyART}\label{fig:time}
\vspace{-0.1in}
\end{figure}

\smallskip
\noindent
\textbf{Efficiency Evaluation (RQ4).} It is important for recommendation tools to respond in a short period of time. According to Figure~\ref{fig:time}, the average API recommendation time of PyART for a query in each project is generally lower than one second, hardly noticeable.

\begin{figure*}[htp]
    \label{sensitive}
    \centering
    \subfigure[Top-1 Accuracy (\%)]{
        \includegraphics[width=2.8in]{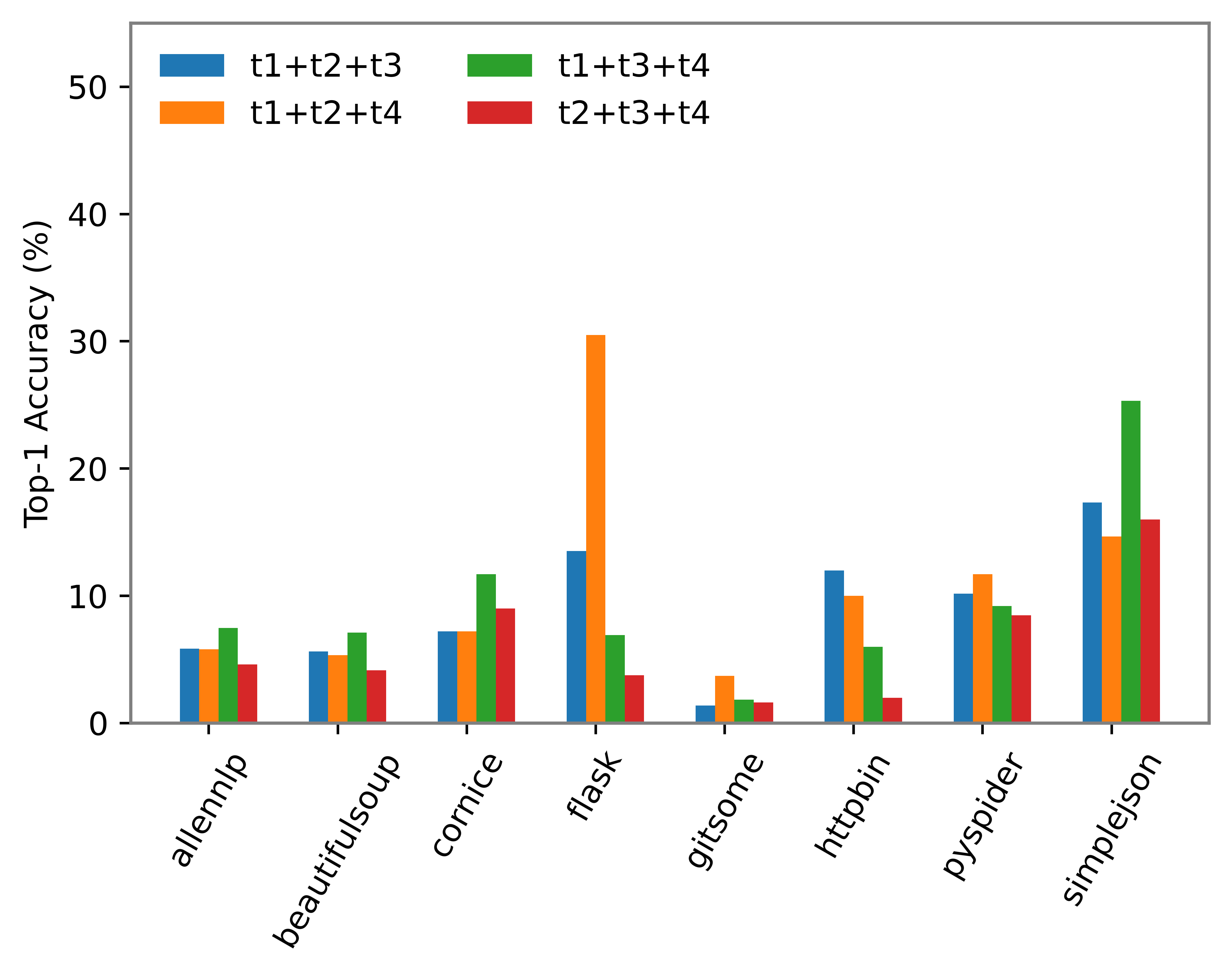}
        \label{label_for_cross_ref_1}
    }
    \subfigure[Top-5 Accuracy (\%)]{
	\includegraphics[width=2.8in]{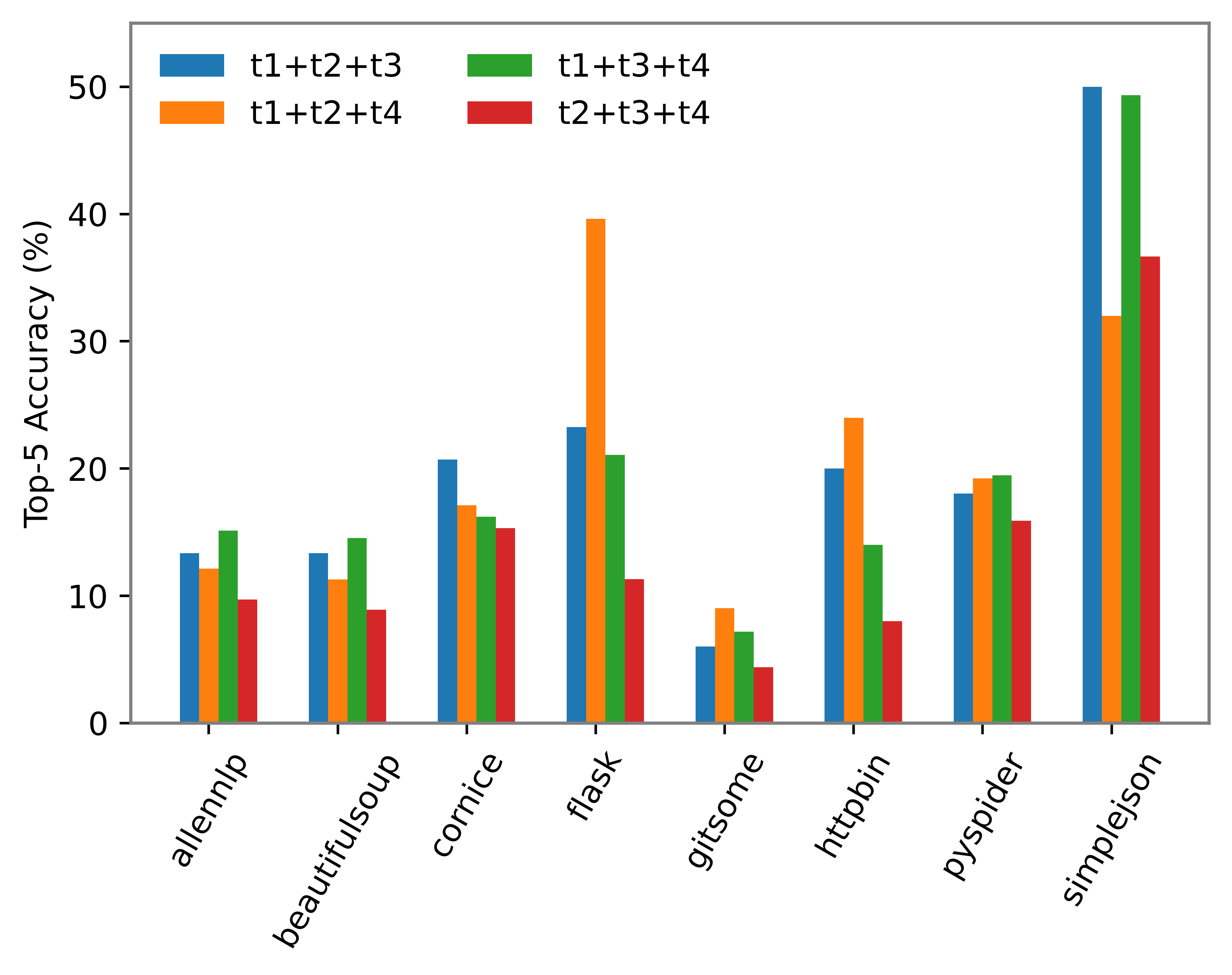}
        \label{label_for_cross_ref_2}
    }   
    \quad
    \subfigure[Top-10 Accuracy (\%)]{
    	\includegraphics[width=2.8in]{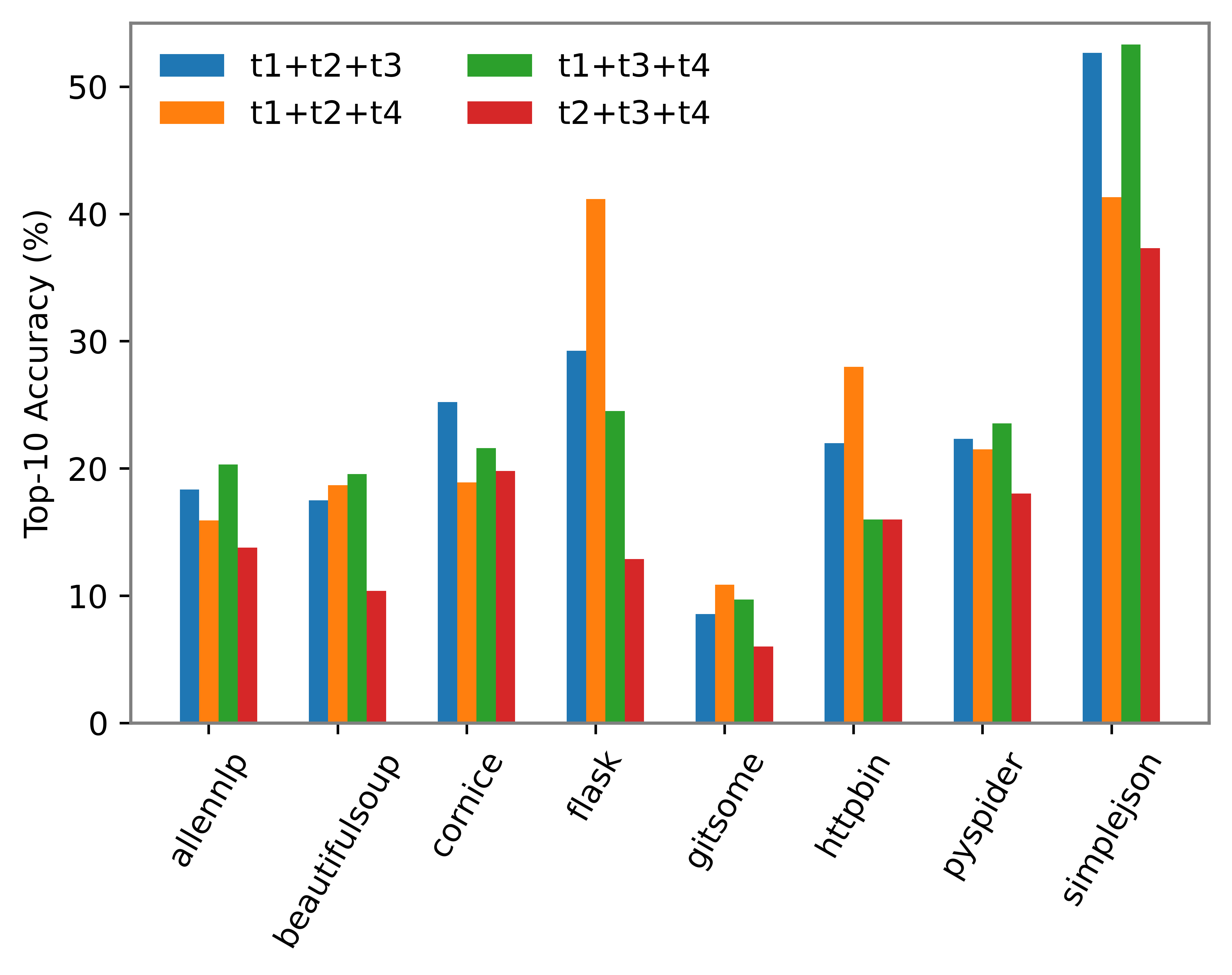}
        \label{label_for_cross_ref_3}
    }
    \subfigure[MRR (\%)]{
	\includegraphics[width=2.8in]{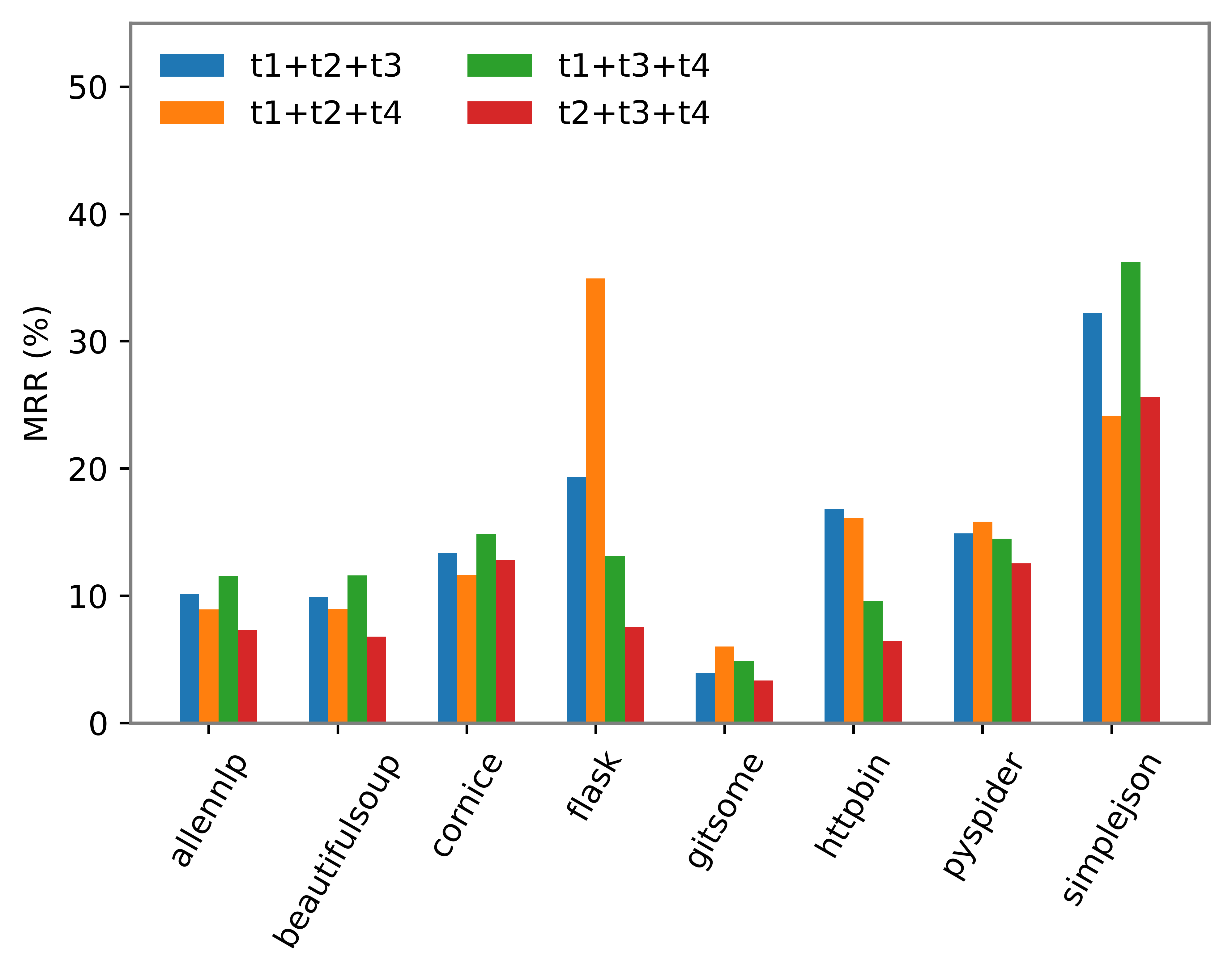}
        \label{label_for_cross_ref_4}
    }
    \caption{Impact of each kind of features on accuracy and MRR
    }\label{fig:sensitive}
\end{figure*}

\smallskip
\noindent
\textbf{Feature Importance Analysis (RQ5).} We evaluate each kind of feature by subtracting one kind at a time to construct a new random forest model. Figure~\ref{fig:sensitive} shows the effects of each kind of feature on top-1 accuracy, top-5 accuracy, top-10 accuracy, and the MRR value. According to the results, we observe that each kind has a non-trivial positive contribution to the accuracy. For 6 out of the 8 projects, accuracy is most heavily affected by the optimistic data-flow feature.


\subsection{Threats to Validity}
\noindent
\textbf{Internal validity.} Our evaluation is performed on a limited set of Python projects. The performance of PyART may vary for different programs. The ground truth of the data flow evaluation and the results of API recommendation by using Intellicode are both provided manually, which may lead to biases. To mitigate such risk, two authors performed the same experiments independently and cross-checked afterwards. Another author was employed to resolve any conflict. The conflict rate is lower than 5\% and all of them were resolved. For validation and reproduction purpose, we release our code and datasets.

\noindent
\textbf{External validity.} There are often run-time errors when using Gumtree and parser on Python programs. This is  different from the case for Java programs. Such errors may lead to biases.


\section{Related Work}\label{related work}
\noindent
\textbf{API recommendation for natural language queries.} 
Researchers use embeddings\cite{yuan2019api,bengio2013representation,coates2011analysis,hamilton2017inductive,li2015word,levy2014neural,lee2020combining,huang2018api,xiong2018automating,sun2019enabling,rahman2016rack} to model and translate among natural language queries, application documentation, and API functionalities. 
Others use graphs for representation and recommendation~\cite{qi2019finding,ling2019graph}. 
These techniques aim to answer queries in natural languages and differ from our real-time scenario.


\noindent
\textbf{Method recommendation based on code corpus and context.} Nguyen et al. \cite{nguyen2015graph} proposed a graph-based  language model Gralan learned from a source code corpus.
Based on the model, an API suggestion engine and an AST-based language model are constructed to make recommendation.
Nguyen et al.
\cite{nguyen2016api} 
proposed to learn from a corpus the regularity between  fine-grained code changes~\cite{negara2014mining,dias2015untangling} 
and APIs, and use that to make recommendations.
Liu et al. \cite{liu2018effective} proposed a method independent of historical
commits 
by ranking the Top-10 recommendations of  Gralan \cite{nguyen2015graph} to achieve better Top-1 accuracy than APIREC \cite{nguyen2016api}. Nguyen et al. \cite{nguyen2019focus} mine open-source software (OSS) repositories\cite{ fitzgerald2006transformation, spinellis2020dataset} to provide developers with API function calls and usage patterns by analyzing how APIs are used in projects similar to the current project using a 3D matrix. 
Although existing works achieve very good results, most of them aim at recommending APIs for static languages such as Java. API recommendation for dynamic languages in real-time poses many new challenges. Sch{\"a}fer et al. \cite{schafer2013effective} proposed an effective smart completion method for JavaScript by combining a static alias analysis\cite{fegade2020scalable,hind2001pointer} enhanced with support for usage-based property inference and a fully automatic dynamic analysis which infers API models based on the framework's test suite.
D'Souza et al. \cite{d2016collective} leveraged API use patterns from open-source repositories to order Python API recommendations. However
it does not focus on addressing the prominent challenges in dynamic languages, such as type dynamics, path sensitivity and third-party libraries. There are some IDEs such as Intellicode \cite{svyatkovskiy2019pythia} that provide real-time recommendations for Python APIs based on artificial intelligence technologies. However, they are often dependent on type information and cannot make accurate recommendations for project-specific APIs.


\section{Conclusion}\label{conclusion}

We propose a novel approach, PyART, to recommending APIs in real-time for Python.
It overcomes the challenges of handling  dynamic language features by extracting so-called optimistic data-flow, which is neither sound nor complete, but resembles the data-flow information humans derive. It also extracts token similarity and token co-occurrences as additional features and encodes them together to numerical vectors.
Random forest is then used to train a model from a corpus of Python projects. Our results show that our technique substantially outperforms a state-of-the-art research prototype and the API recommendation technique in a mainstream Python IDE.


\section*{Acknowledgment}

We thank the anonymous reviewers for their constructive comments. This research was supported, in part by NSFC 61832009, Jiangsu Postgraduate Innovation Program KYCX20\_0041, Cooperation Fund of  Huawei-Nanjing University Next Generation Programming Innovation Lab (No. YBN2019105178SW11), NSF 1901242 and 1910300, ONR N000141712045, N000141410468 and N000141712947. Any opinions, findings, and conclusion in this paper are those of the authors only and do not necessarily reflect the views of our sponsors.

\vspace{1.8cm}

\bibliographystyle{IEEEtran}
\bibliography{IEEEabrv,sample-base}

\end{document}